\documentclass[sigconf]{acmart}

\usepackage[utf8]{inputenc} 
\usepackage[T1]{fontenc}    
\usepackage{hyperref}       
\usepackage{url}            
\usepackage{booktabs}       
\usepackage{nicefrac}       
\usepackage{microtype}      
\usepackage{xcolor}         
\usepackage{graphicx}
\usepackage{bm}
\usepackage{amsmath}
\usepackage{soul}  
\usepackage{xcolor}  
\usepackage{algorithm}
\usepackage{algpseudocode}

\newcommand{\highlight}[1]{#1} 

\title{HRTF Estimation in the Wild}

\author{Vivek Jayaram}
\email{vjayaram@cs.washington.edu}
\affiliation{%
  \institution{University of Washington}
  \city{Seattle}
  \state{WA}
  \country{USA}
}

\author{Ira Kemelmacher-Shlizerman}
\email{kemelmi@cs.washington.edu}
\affiliation{%
  \institution{University of Washington}
  \city{Seattle}
  \state{WA}
  \country{USA}
}

\author{Steven M. Seitz}
\email{seitz@cs.washington.edu}
\affiliation{%
  \institution{University of Washington}
  \city{Seattle}
  \state{WA}
  \country{USA}
}

\copyrightyear{2023}
\acmYear{2023}
\setcopyright{rightsretained}
\acmConference[UIST '23]{The 36th Annual ACM Symposium on User Interface
Software and Technology}{October 29-November 1, 2023}{San Francisco, CA,
USA}
\acmBooktitle{The 36th Annual ACM Symposium on User Interface Software and
Technology (UIST '23), October 29-November 1, 2023, San Francisco, CA, USA}
\acmDOI{10.1145/3586183.3606782}
\acmISBN{979-8-4007-0132-0/23/10}

\begin{document}

\begin{abstract}
Head Related Transfer Functions (HRTFs) play a crucial role in creating immersive spatial audio experiences. However, HRTFs differ significantly from person to person, and traditional methods for estimating personalized HRTFs are expensive, time-consuming, and require specialized equipment. We imagine a world where your personalized HRTF can be determined by capturing data through earbuds in everyday environments. In this paper, we propose a novel approach for deriving personalized HRTFs that only relies on in-the-wild binaural recordings and head tracking data. By analyzing how sounds change as the user rotates their head through different environments with different noise sources, we can accurately estimate their personalized HRTF. Our results show that our predicted HRTFs closely match ground-truth HRTFs measured in an anechoic chamber. Furthermore, listening studies demonstrate that our personalized HRTFs significantly improve sound localization and reduce front-back confusion in virtual environments. Our approach offers an efficient and accessible method for deriving personalized HRTFs and has the potential to greatly improve spatial audio experiences. Video demos and code can be found on our project website.\footnote{\url{https://grail.cs.washington.edu/projects/hrtf-in-the-wild/}}

\end{abstract}



\begin{teaserfigure}
  \centering
  \includegraphics[width=0.8\textwidth]{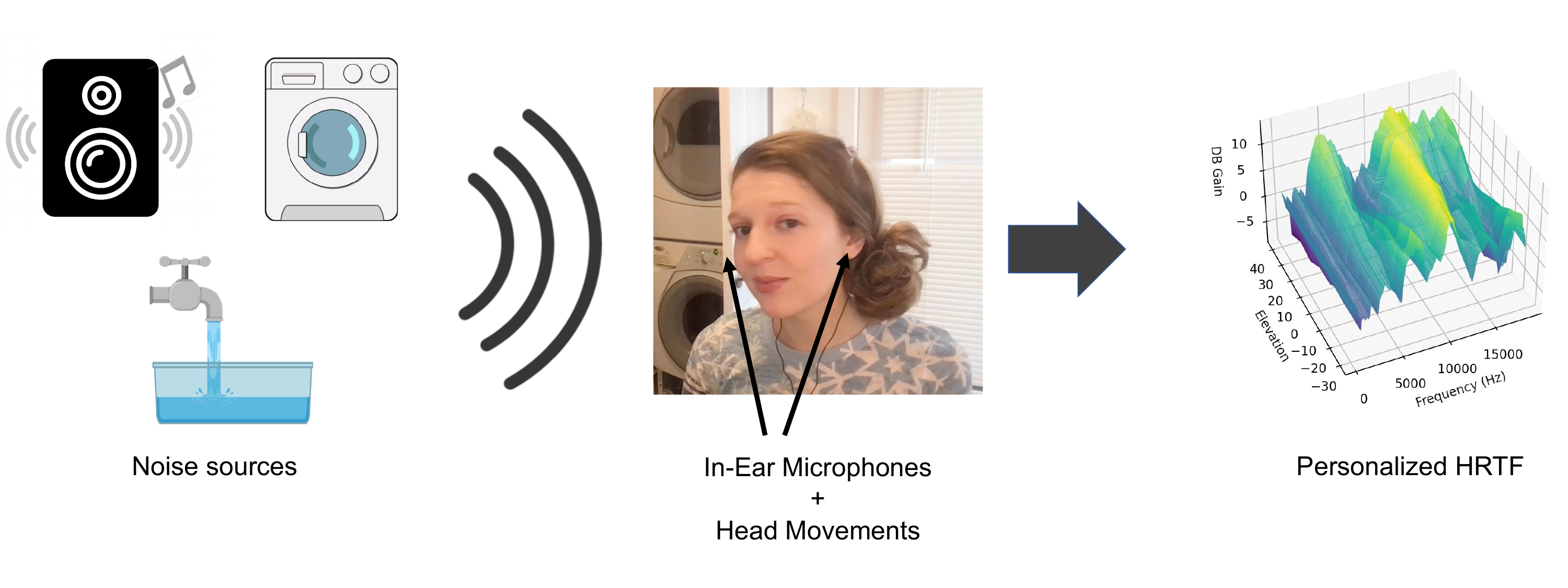}
  \vspace{-10pt}
  \captionsetup{width=0.8\textwidth}
  \caption{Our method uses binaural recordings of everyday noises along with head tracking information to create a personalized HRTF for the listener.}
  \label{fig:teaser}
  \vspace{15pt}
\end{teaserfigure}
\maketitle

\section{Intro}

Spatial audio is an important aspect of many audio applications, including virtual and augmented reality, gaming, music, and audio for film and television. The fundamental challenge of spatial audio is to create the perception that sound is coming from any location in space, even though the sound is played back through headphones. Humans are remarkably good at perceiving the location of incoming sounds in the real world, with as little as $3.5^{\circ}$ error even in noisy environments \cite{makous1990two}. This ability is achieved through the Head Related Transfer Function (HRTF), which is the direction-dependent filtering of sound by the head, ears, and torso. By using a listener's HRTF to render virtual sounds, it is possible to create an immersive audio experience that simulates sound coming from any position in 3D space.  The HRTF is comprised of two components: interaural time differences (ITD) and interaural level differences (ILD). While both components are important for accurate spatial localization, this paper focuses on the frequency dependent ILDs, also called spectral features which describe the different frequencies arriving at each ear. These are more easily obtainable from in-the-wild recordings and have been shown to be more important for HRTF personalization compared to ITDs \cite{wenzel1993localization}.

A key problem is that HRTFs vary significantly from person to person, and using a personalized HRTF is necessary to create high fidelity spatial audio. This is because using someone else's HRTF or a generic HRTF will lead to localization errors and an unpleasant listening experience \cite{wenzel1993localization, middlebrooks1999individual}. Despite its importance, accurately measuring an individual HRTF is a difficult task. This is due to the fact that HRTF is a complex, dynamic phenomenon that is affected by a variety of factors, including an individual's ear shape, head size, and general anatomy. Traditional methods require the listener to sit in an anechoic chamber while sine-sweeps are played from all possible angles. Other methods involve taking complex 3D scans of the head and ears along with anatomical measurements. This problem of personalized spatial audio has also received increasing attention from companies, such as Apple, Sony, and Logitech, which have recently developed methods to create personalized HRTFs through head scans, imaging, and user feedback \cite{applepersonalized, sonypersonalized, logitechpersonalized}. Despite these advances, achieving high-fidelity spatial audio remains an ongoing challenge and an active area of research and development.

We are particularly motivated by the rapid proliferation of earbuds systems, with 100 million AirPods
sold in 2020 alone \cite{airpodssales}. These systems typically contain a microphone in each ear as well as a head tracking IMU, making them ideal for capturing personalized HRTFs. As more and more people use earbuds, we envision a future where collecting data for personalized HRTFs is as simple as wearing earbuds and moving around in different environments. By analyzing the changes in sound arriving at the listener's ears over time, we can infer their personalized HRTF and use it across a wide range of spatial audio applications. This approach has the potential to be more efficient and less burdensome than traditional methods that require 3D scans or anthropometric measurements.

As a step towards that, in this paper we present a method for measuring individualized HRTFs that leverages environmental sounds recorded by the listener in everyday settings. Our approach is designed for scenarios where there is a single stationary noise source, and we demonstrate its effectiveness using a wide range of noise sources such as music, home appliances, and outdoor sounds. By analyzing the recorded sounds, we can extract features that are specific to the listener's HRTF and use them to construct a personalized HRTF. Our method utilizes machine learning along with synthetic and real training data in order to predict the frequency-dependent filtering of a subject from natural recordings. 

Because binaural microphone data and head tracking information is not available through the public APIs of current earbud systems, we built our own physical system to resemble the data available from these earbuds. There have already been some commercial headphones that enable binaural recordings for developers \cite{Chatterjee_2022, samsungbuds} so we expect to see this data becoming more accessible to developers over time.

To validate our approach, we conducted user studies with real listeners and show three key experimental results. First, our predicted HRTFs closely match the ground truth HRTF recorded in an anechoic chamber. Second, our HRTFs significantly improve the sound localization accuracy of users in a virtual auditory display when compared to a generic HRTF. Third, our HRTFs greatly reduce front-back confusion when used to render sounds. Overall, our proposed method of measuring individualized HRTFs in-the-wild has the potential to offer a more efficient and less burdensome alternative to traditional methods, and we hope that it will inspire further research and development in this field.
\section{Related Works}
\begin{figure*}
    \centering
  \includegraphics[width=0.95\textwidth]{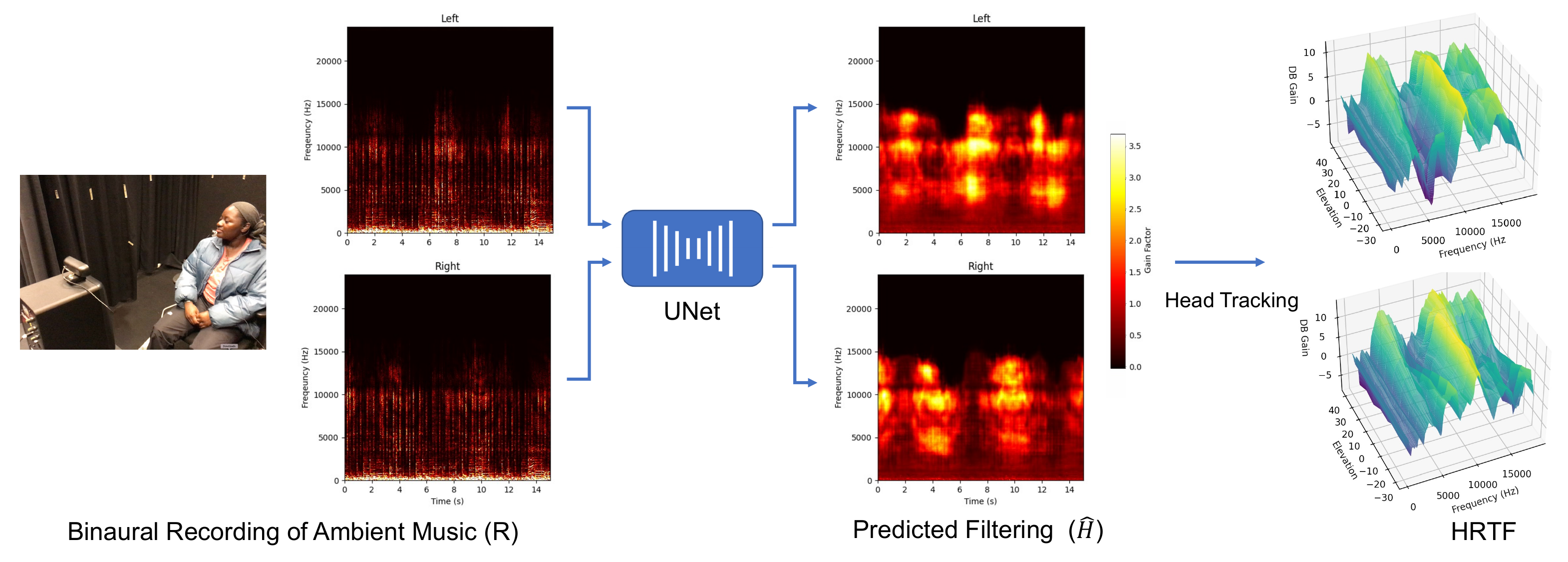}
  \caption{An overview of our method. We use binaural recordings of in-the-wild sounds to predict the filtering from the HRTF at each time step. We then use the head tracking data to map this predicted filtering to the user's location dependent HRTF.}
  \label{fig:method}
\end{figure*}

Traditional methods of measuring an individual's HRTF involve dense acoustic measurement in an anechoic chamber \cite{sridhar2017database, algazi2001cipic, moller1995head, watanabe2014dataset, southampton}. The listener is positioned in the center of the chamber and provided with in-ear microphones, while a series of loudspeakers are arranged in a spherical array to cover all possible azimuth and elevation angles. A reference signal, such as an exponential sine sweep, is played one at a time from each speaker location, and the resulting sound wave that reaches the ear is compared to the reference signal to determine the HRTF. In some cases, the speakers are placed on an arc and rotated around the subject to reduce the number of required speakers. While these approaches are accurate and provide high fidelity HRTFs, they are time-consuming and resource-intensive and require the listener to come to a specialized lab for measurement.

In order to speed up this process, other methods have been proposed that only use a single loud speaker \cite{reijniers2017diy, headmovements, fast2dheadmovements}. In these works, a reference signal is played from a stationary loudspeaker while the subject rotates their head through different directions under the measurement of an IMU or other head tracking device. This has the possibility to greatly simplify the HRTF measurement process, and our method builds on this idea of measuring the listener's motion relative to the noise source.  


Acoustic simulations on 3D scans of individuals represent another broad category of HRTF estimation methods. For example, the algorithms described in \cite{ziegelwanger2015mesh2hrtf, huttunen2014rapid, huttunen2007simulation, pHRTF, mesharmmesh} use 3D mesh data with boundary-element methods to simulate the diffraction of sound waves through the head and ear. It has also been shown that HRTFs can be calculated directly from a point cloud of the head, which is slightly easier to obtain than a 3D mesh \cite{sridhar2017method}. Although not published, the method released by Apple \cite{applepersonalized} uses the depth sensor to create a 3D scan of the head. 3D scan based methods are more accessible than traditional acoustic methods, but still suffer from several drawbacks. For one, they rely on an accurate 3D mesh which can be hard to obtain, either requiring a depth sensor or many images. In addition, 3D scans and imaging raise more privacy concerns for users compared to audio only methods.

It is also possible to estimate the HRTF directly from anthropometric measurements, given the availability of large HRTF datasets with associated head and ear measurements \cite{algazi2001cipic, watanabe2014dataset}. The works in \cite{zotkin2002customizable, anthropometric} show positive results when selecting the HRTF with the closest anthropometric measurements to a new user. Other works, \cite{hu2006head, zhao2022magnitude, chun2017deep, chen2019autoencoding}, use regression methods or deep learning to predict HRTF features from these anthropometric measurements, including works like \cite{zhiphoto, zhao2022magnitude, mohancomputervision} that use images of the ears along with anthropometric measurements. Anthropometric methods face the same challenges as 3D scan-based methods, as obtaining accurate measurements is difficult and can be time-consuming.

Recently, methods have been proposed to measure HRTFs acoustically in less controlled environments. The method in \cite{diepold2010hrtf} proposed measuring the HRTF from everyday recordings, but uses a third microphone in the room as a way to record the clean reference signal. Another method \cite{zandi2022individualizing} allows the user to play sine sweeps from their smartphone, but requires capturing this signal from at least 60 unique locations around the head. Similarly the method in \cite{earables} asks the user to play predefined sounds from their smartphone while they move the phone around their head. Finally, the method in \cite{yamamoto2017fully} allows users to answer pairwise comparison questions in a listening study to determine the best HRTF.

In contrast to these methods, our method has a few key advantages. First, we don't require an anechoic chamber or specialized speakers. Second, we don't require any 3D scans or imaging of the head. Finally, the recordings are collected passively from sound sources in the environment. \highlight{In our method, the user is in an everyday environment and we capture their natural head movements after an initial calibration step. This is meant to be less cumbersome than existing methods that involve answering questions, moving a smartphone around, or take detailed head measurements and scans.}

\section{Method}

Suppose that a listener is wearing earbuds which contain a microphone in each ear as well as a head tracking IMU. The listener may be in the presence of a sound source $s$, and the microphones at each ear will pick up the binaural recording given by $r_l$ and $r_r$ for left and right respectively. We also use the 3DoF head rotation: $\bm{\theta}_h(t)$ which describes the head rotation at any given moment in time $t$ This data is available from recent airpod devices \cite{airpodspecs}, or could be captured through the webcam. Our goal is to learn the HRTF solely from $r_l$, $r_r$, and $\bm{\theta}_h(t)$. We use the uppercase notation $S$, $R_l$, and $R_r$ to refer to the time-frequency representation of the audio, and the lowercase to refer to the waveform representation. Similarly we use $H$ to denote the filtering function imposed by the listener during a recording, and the time domain version, the head related impulse response (HRIR) is written as $h$. In this work, we limit the method to scenarios with a single stationary sound source, and its position relative to the head is written as $\theta_s(t)$, and $\varphi_s(t)$.


Under the simplest assumptions, the captured audio is a convolution between the HRIR and the original source. For example, the recorded audio $r$ can be written as
\begin{equation}
r = h \ast s
\end{equation}
In the frequency domain, this is
\begin{equation}
R = H \cdot S
\end{equation}
or equivalently
\begin{equation}
H = R / S
\end{equation}

Furthermore, the recording may include multi-path signals and other ambient noise, denoted as $\epsilon$, which add ambiguity to the scenario. Breaking this down by left and right separately we get

\begin{equation}
H_l = (R_l - \epsilon_l) / S
\end{equation}
\begin{equation}
H_r = (R_r - \epsilon_r) / S
\end{equation}

As we can see, this is a highly underdetermined problem, since we do not have access to the original sound source $S$ or multipath contributions $\epsilon$, only the captured recordings $R_l$ and $R_r$. Therefore, at any given moment, we would not know whether a given frequency was modified by the listener's HRTF or by the emitting sound source.

Our goal is then to predict $H_l$ and $H_r$ from $R_l$ and $R_r$ without access to the actual ground truth source S. We can then use $H_l$ and $H_r$ along with the head tracking information to create a listener specific HRTF, which is a function of the source location and frequency: $HRTF(\theta_s, \varphi_s, f)$

\subsection{Deep Network}

To solve this problem, we can use the fact that most sound sources have a repeated or predictable frequency distribution over time which can be learned. Furthermore, the recording from both ears together provide clues towards the relative filtering at each ear.  We frame this problem as a supervised learning problem and use a deep network for this prediction task. Deep networks can learn the underlying structure of sounds such as speech and various noise sources, solving one of the ambiguities. Secondly, these networks can use the temporal information of the source along with the data captured at both ears to predict which frequencies are being modified by the HRTF instead of by the sound source or multipaths.

Our network is a modification of the Unet Convolutional Neural Network \cite{unet} \highlight{with an initial convolutional block comprising 32 features, and composed of of 4 downsampling and 4 upsampling convolutional blocks}. $R_l$ and $R_r$ are produced using the magnitude of a short-time Fourier transform of the captured audio. They are concatenated channel wise, and feed through the network to produce 2 channels of output of the same dimension. The output represents the predicted level change at a certain frequency due to the HRTF as a scalar factor. We found that learning the filtering function as multiplicative gain was easier than dB due to the fact that cutting out a frequency would require learning a dB gain of $-\infty$. Training details are described in Section \ref{sec:training}.

\subsection{Source Localization and Head Tracking}
Head tracking through an IMU or camera can provide the 3DoF rotation angle of the head. However, because the sound source may not be located directly in front of the user, it is also necessary to know the location of the signal relative to the user. In our system, we require an initial localization input from the user. They are asked to point their head directly towards the sound source (or directly away for sounds coming from behind). They then press a button which allows the system to record the initial location of the sound source, $\theta_s(0)$ and $\varphi_s(0)$. During the rest of the recording process, the rotation matrix of the head orientation can be applied to the initial source location to give the relative position of the sound source at that time, $\theta_s(t)$ and $\varphi_s(t)$.

It may also be possible to infer the initial source location using localization algorithms, but we leave that as future work as the manual localization by the user is quick and very accurate.

\subsection{HRTF Estimation from Aggregated Results}
By aggregating predictions across many recordings with different sound sources and head rotations, we can obtain a more accurate and full representation of the listener's HRTF. Let $F$ be the number of frequency bins in the spectrogram representation. For each binaural recording $R \in \mathbb{R}^{2 \times T \times F}$, we use a deep network to predict the filtering function $\hat{H} \in \mathbb{R}^{2 \times T \times F} = \text{UNet}(R)$.

To build a model of the listener's HRTF, we first initialize an empty HRTF for all source locations and frequencies. Then, we use the UNet to predict how the listener's HRTF filtered the sound source for each recording. If the entirety of a recording contains minimal energy at a given frequency, we assume that this frequency was absent from the source signal and do not use it. Finally, we use the known relative location of the source over time to create a HRTF prediction for each location-frequency bin. \highlight{Our method does not explicitly solve for directions with no data, but in such scenarios, we could use HRTF extrapolation/interpolation methods which have shown good results when we only have a sparse HRTF} \cite{ben-hur2020localization, ito2022headrelated}

\highlight{Across time steps, the predicted filtering function for the same location-frequency bin may vary due to the changes in the underlying sound source, reverb, or other effects not modeled such as doppler effects. Because of this, we average the predicted HRTF values at each location-frequency bin to obtain the listener's HRTF magnitude at each location and frequency. One of the advantages of our method is that over time, we can collect more and more information about the HRTF and use that to produce a better estimate. We explored both the mean and median and found that the mean worked better. }

To obtain the head-related impulse response (HRIR) for use in spatial audio applications, we also need the phase information which describes the interaural time differences (ITDs). We use ITDs from a generic HRTF and apply inverse fast Fourier transform (IFFT) to obtain the HRIR. \highlight{Although some previous works in similar domains} \cite{richard2022deep, filteredNoiseRIR} \highlight{predict the phase as well as the magnitude of the impulse response, we found that phase was much harder to predict in a reverberant environment due to multipath effects. At many frequencies, the captured phase was completely different from the actual ITD phase due to multipath interference. Our user studies also showed that generic ITDs still produced a strong ability to localize sounds}. The full algorithm is described in Algorithm \ref{hrtfcreation}.

\begin{algorithm}[t]
\caption{Create HRTF, HRIRs from Binaural Recordings}\label{hrtfcreation}
\begin{algorithmic}[1]
    \For{$\forall \theta, \forall \varphi, \forall f$}
    \State $\hat{HRTF}(\theta, \phi, f) \gets []$ \Comment{Initialize empty HRTF}
    \EndFor
    \For{$R \in \texttt{Recordings}$}
        \State $\hat{H} \gets \text{UNet}(R)$ \Comment{Network inference}
        \For{$t \in 0..T$, $f \in 0..F$}
                \If{$R(f).\texttt{mean()} > \epsilon$}
                \State $\hat{HRTF}(\theta_s(t), \phi_s(t), f).\text{append}(\hat{H}(t, f))$
                \EndIf
        \EndFor
    \EndFor
  \For{$\forall \theta, \forall \varphi, \forall f$}  \Comment{Use phase from generic HRTF}
  \State $| \hat{HRTF}(\theta, \phi, f) | \gets \hat{HRTF}(\theta, \phi, f).\texttt{mean()}$
  \State $\angle \hat{HRTF}(\theta, \phi, f) \gets \angle HRTF_{\texttt{generic}}(\theta, \phi, f)$
  \EndFor
\State $HRIR(\theta, \varphi) = \texttt{iFFT}(HRTF(\theta, \varphi))$
\end{algorithmic}
\end{algorithm}

\subsection{System Implementation}
Our method is general and designed to work with any device that supports binaural recordings and head tracking. This could include earbuds, VR headsets, or smart glasses. However, with the exception of certain headsets paired with certain phones \cite{samsungbuds, airpodscmmotion}, these devices do not currently expose the required functionality to third-party developers. We therefore built our own physical system with commercially available hardware.

For the binaural recordings, we used the Sound Professionals SP-TFB-2 in-ear Binaural Microphones \cite{binauralmics}. These wired headphones are capable of capturing frequencies up to 20kHz. \highlight{It's noteworthy that our microphones, unlike those used in numerous previous studies such as} \cite{sridhar2017database,sridhar2017method, algazi2001cipic}\highlight{, are positioned at the entrance of the ear canal rather than fully blocking it. Our research demonstrates that it is feasible to generate an accurate HRTF even without perfect microphone placement. Extending this methodology to commercial earbuds would require re-training with data captured using those specific headphones to learn their unique transfer function.}

For head tracking, we used the face pose detector provided by the Google MediaPipe Library \cite{googlemediapipe}. This algorithm uses a forward facing webcam to detect the 3DoF head position, and is based on BlazeFace \cite{bazarevsky2019blazeface} and AttentionMesh \cite{grishchenko2020attention}. The head tracking runs in less than $10\texttt{ms}$ on a Macbook pro, and we use a HRTF with bin size $\theta = 5^{\circ}$ and  $\varphi = 5^{\circ}$. This means that as long as the user is not rotating their head faster than $\sim300^{\circ} / s$, the head tracking will assign the sound to the correct HRTF bin.

\begin{figure}[htb]
  \centering
  \includegraphics[width=0.8 \linewidth]{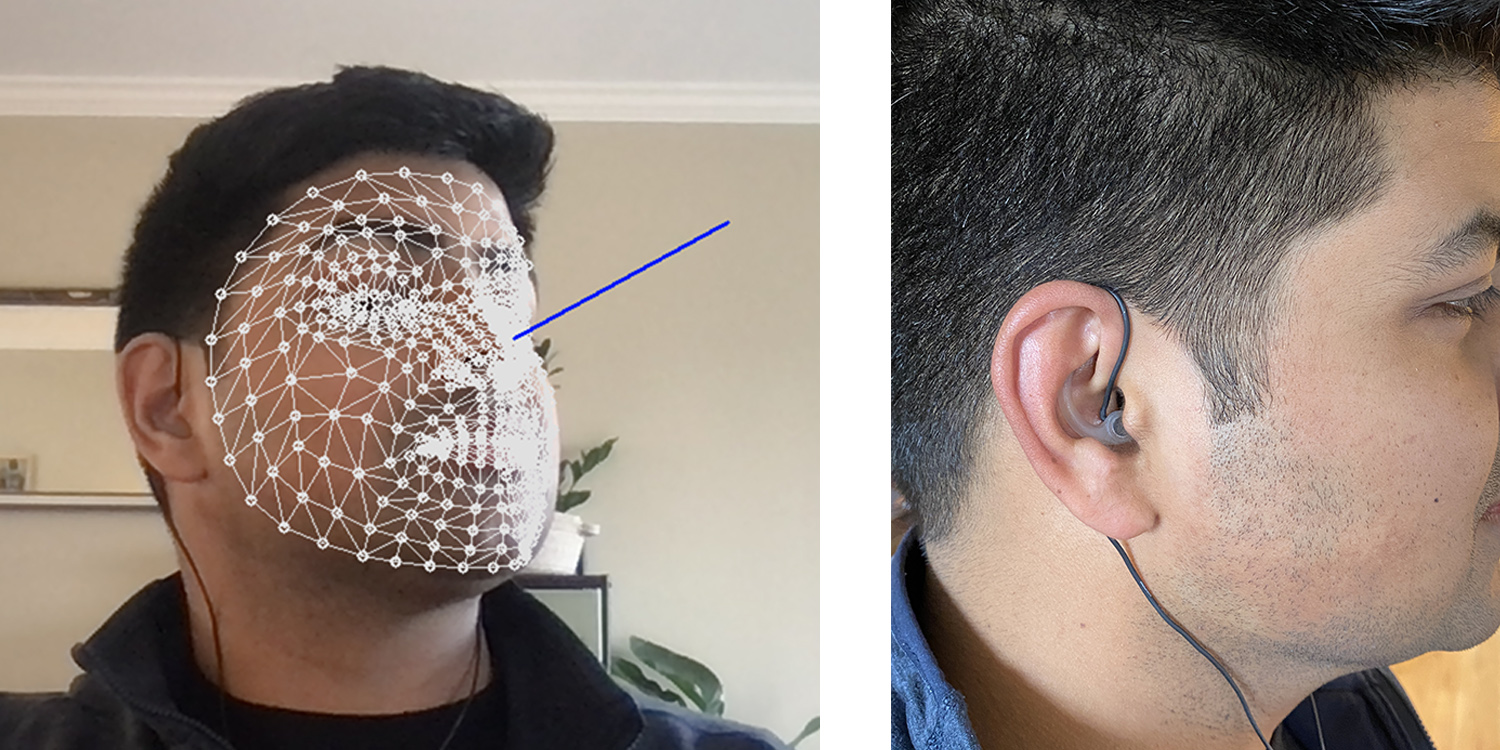}
  \caption{Left: Our head tracking implementation uses the webcam to determine the 3DoF head rotation during recording. The normal vector is drawn in blue to help visualize the direction the head is pointing. Right: An image of the binaural microphone used in our implementation. The microphone sits near the ear canal.}
  \label{fig:head_tracking}
\end{figure}

\section{Data and Training}\label{sec:training}
\begin{figure*}
    \centering
  \includegraphics[width=0.95\textwidth]{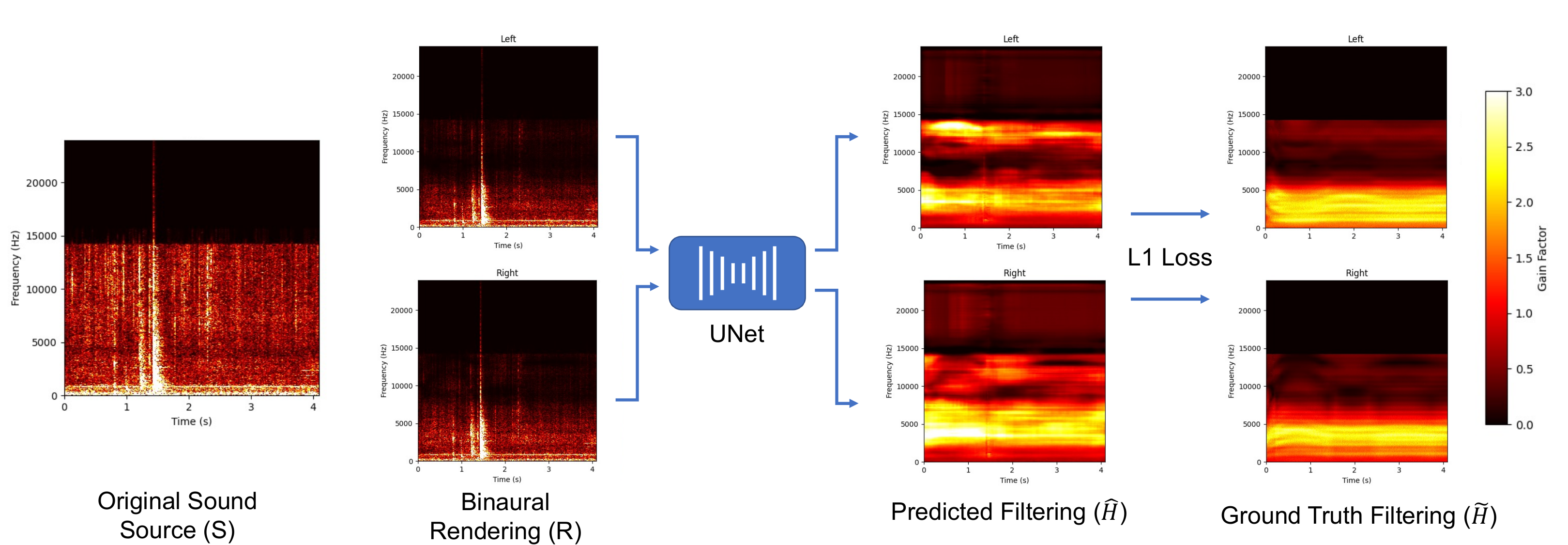}
  \caption{The process for training the network. We use create binaural renderings of a sound source with simulated multi-path environments. We then use the ground truth filtering of the HRTF to train the network with an L1 loss between the predicted filtering, $\hat{H}$ an the ground truth filtering $\tilde{H}$. The real training data is used in an identical way except $R$ is a binaural recording, not a binaural rendering, and we don't have access to the original sound source $S$.}
  \label{fig:method}
\end{figure*}
To train our network, we adopt an approach that combines synthetic and real data. We begin with large amounts of synthetically rendered data, which enables us to learn from a wide range of noise types and simulated environments, including multi-path scenarios. However, such data does not capture all nuances of real-world audio and fails to generalize completely to actual recordings. To address this, we incorporate real data, which is more time-consuming to collect but provides more effective training for the network. By leveraging both sources of data, we are able to benefit from the strengths of each approach. This mix of synthetic and real training data has been explored in previous works as well \cite{Chatterjee_2022, jenrungrot2020cone, seib2020mixing}. Below, we describe the two data sources in more detail.

\subsection{Synthetic Training Data}
We first train the network on synthetically rendered spatial data. For HRTFs, we use the RIEC dataset \cite{majdak2013spatially}, which contains 109 HRTFs measured in an anechoic chamber for different individuals. This was split into a training set of 64 and a test set of 45 HRTFs. To render sounds, we use the Steam Audio C++ API which allows realistic sound rendering for moving sources with custom HRTFs and multi-path environments. 

The noise sources come from the  WHAM! dataset \cite{wham} and AudioSet dataset \cite{audioset}. These datasets contain a wide variety of noise sources such as music, speech, appliances, and machinery. Sound sources without sufficient frequency ranges (requiring a minimal energy up to at least 5khz) and without sufficient regularity (e.g. impact only sounds) were filtered out. Some example of sound categories that were removed included chewing, clapping, snapping, and whistling. Some of the most effective noise sources included water, kitchen appliances, pop music, and machinery. \highlight{For both datasets, a 80/20 train/test split was maintained. None of the audio samples used for training the network were used during any of the synthetic or real evaluations}

Each generated recording was 3s long and created by placing the sound source at a random azimuth and random elevation 1.5m away from the listener while the listener moved their head in a random direction at a random speed. Multipaths were simulated by create walls at distances between 2 and 10 meters from the listener with RT60 values from \highlight{0.4 to 0.9 seconds}. For each recording we also obtained the ground truth filtering at the source locations to use as a training label, $\tilde{H}$. The train set contains 10,000 generated examples, and the test set contains 1,000 examples.

\subsection{Real Training Data and Anechoic HRTFs}\label{sec:anechoic}

Although large amounts of synthetic data can be easily collected, a network solely trained on synthetic data does not perform well in real-world scenarios. To address this issue, we augment the training data with in-the-wild recordings that more closely resemble the acoustic environments and sounds that would be encountered by users during inference. The main challenge is that the ground-truth filtering function is required as a training supervised label for the model. In order to generate these labels, we measure the ground-truth HRTFs of the subjects in an anechoic chamber. These HRTFs can also be used as a baseline to evaluate the in-the-wild inferred HRTFs.

Our anechoic HRTF measurement procedure is most similar to the method described in \cite{headmovements}. Subjects are seated in an anechoic chamber while a single loudspeaker emits a reference signal. They are instructed to move their head slowly to cover a broad range of azimuth and elevation angles. Unlike \cite{headmovements}, we place the speaker at 3 different elevation angles when capturing the ground-truth HRTF. This better captures the filtering effects of the torso at different sound elevations which are not captured by simply rotating the head up and down with respect to a single speaker location. Furthermore, we use a broadband Gaussian noise signal instead of a sine sweep, as we only care about the frequency-dependent level differences and not the ITDs. This allows us to capture the filtering across all frequencies at each time step. The speaker used is the KRK Classic 5 Studio Monitor which contains 2 drivers. To account for an imperfect speaker response, a reference signal $\tilde{S}$ is first recorded. The ground truth filtering function is then obtained by dividing the recording $R$ by $\tilde{S}$. This also has the effect of cancelling out any frequency response imposed by the microphones as both $R$ and $\tilde{S}$ contain the same microphone response. A full discussion of speaker and microphone compensation is provided in \cite{langendijk}.

After collecting the anechoic HRTFs, we generated real training data for the neural network by having 2 subjects listen to 1 hour of noise sources, \highlight{from the training partition of our audio datasets}, played back through the loudspeaker in regular environments. The speaker location was known, and the training label $\tilde{H}$ could be generated from the anechoic HRTFs.

\subsection{Training Details}
All recordings were captured at 48kHz sample rate. Each training example contained 3s of binaural audio, and mini-batch size 32 was used. The STFT was conducted with a window size of 2048. Training occured on a Nvidia Titan Xp GPU and took approximately 10 hours for 100 epochs of training. Data augmentation techniques included random left-right flip, random volume changes, and the addition of random noise. \highlight{Samples from the real and synthetic dataset were randomly sampled with equal probability}

\begin{figure*}[t!]
    \centering
  \includegraphics[width=\textwidth]{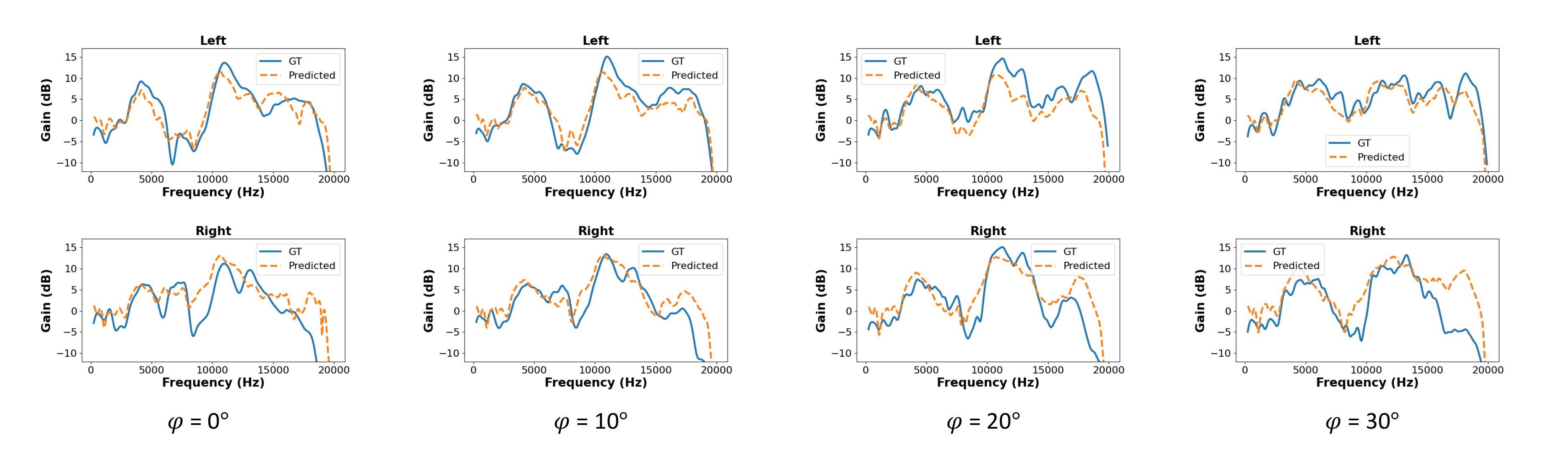}
  \vspace{-25pt}
  \caption{We plot the ground-truth HRTF and predicted HRTF for a given test subject for $\theta=0^\circ$ and 4 elevations. The HRTF that we create for the user closely matches the ground truth, even though the magnitude of some notches and peaks may not be exactly correct.}
  \label{fig:comparison}
\end{figure*}



\section{Results}

We evaluate the effectiveness of our method through a user study, and present three key results to show the strength of the method. First, we show that our predicted HRTFs closely match the ground-truth HRTFs. Second, we demonstrate that our HRTFs improve localization by listeners in a virtual environment. Finally, we show that our HRTFs significantly reduce front-back confusion with rendered sounds.

\subsection{User Study and In-the-Wild HRTF}\label{sec:userstudy}
8 individuals with regular hearing abilities (4 male, 4 female, mean age 28) participated in the user study. First, we measured their ground-truth HRTF in an anechoic chamber as described in Section \ref{sec:anechoic}. Next, we used our in-the-wild method to measure their HRTF in a regular environment. The subjects were in a normal sized reverberant room, that was not particularly quiet. The background noise in the room was measured to be around $50 \texttt{dB}$ due to electric hum and other noises. Next, a variety of noise types were played from a loud speaker in the room. This included music, running water, kitchen appliances, and other sounds from the \highlight{test partition of the} WHAM! and AudioSet datasets. The speaker was placed at 3 elevations and a variety of azimuth angles relative to the listener at distances that varied from $1-3\texttt{m}$. The listener was instructed to rotate their head through a normal range of angles as they listened to the audio sounds. In total, roughly 15 minutes of audio were captured per user across all the locations.

\subsection{Comparison with Ground-Truth HRTF}
The first metric we use to evaluate the correctness of our HRTFs is the agreement with the ground-truth HRTF. A visual comparison between the two is shown in Figure \ref{fig:comparison} which plots the results for a given subject at four consecutive elevations and $\theta=0^\circ$. To evaluate the similarity quantitatively, at every azimuth and elevation, we compute the Log-Spectral Distortion (LSD) in dB which is given by
\begin{equation}
LSD(\hat{H}, \tilde{H}) = \sqrt{\frac{1}{F}\sum_{f=1}^{F}\bigg(20\log_{10}\bigg|\frac{\tilde{H}(k)}{\hat{H}(k}\bigg|\bigg)^{2}}
\end{equation}

\begin{table}
  \label{tab:comparisons}
  \begin{tabular}{cc}
    \toprule
    Method & LSD (dB) \\
    \midrule
    Random RIEC Subject & 8.23 \\
    Generic HRTF & 7.32 \\
    Zandi et. al \cite{zandi2022individualizing} & 4.5 \\
    \textbf{Ours} & \textbf{4.38} \\
    Hu et. al \cite{hu2006head} & 3.5 \\
    \bottomrule
  \end{tabular}
  \vspace{10pt}
  \caption{Log-spectral distortion between ground-truth HRTF and the output HRTF for several methods. We note that the method in \cite{hu2006head} requires additional physical measurements and the method in \cite{zandi2022individualizing} requires significantly more active input from the user.}
  \vspace{-20pt}
\end{table}

We then report the median value across all azimuth and elevations in table 1. Our method is compared with several other methods as well. For the random method, we average the LSD when comparing the ground-truth HRTF with all other HRTFs in the RIEC database. For a generic HRTF, we used the KEMAR HRTF \cite{gardner1994hrft} which contains measurements for a dummy head commonly used as a generic HRTF model. We also share the results reported in \cite{hu2006head} and \cite{zandi2022individualizing}. It's important to note that the method in \cite{zandi2022individualizing} achieves it's reported results when the HRTF is measured at 1138 unique locations done actively by the user, and the method in \cite{hu2006head} requires detailed anthropometric measurements of the ear, head, and torso.

\subsection{Localization in a Virtual Auditory Display}

To evaluate the effectiveness of our HRTFs in spatial audio applications, we created a virtual auditory display where sounds could be rendered spatially with dynamic head tracking. Our method aimed to replicate the experimental method described in \cite{ben-hur2020localization}. A sound reproduction system was implemented in Unity and Steam Audio where a virtual sound was placed at an arbitrary location and played back to the listener through headphones. The listener could then move their head, with the sound location adjusting accordingly based on head tracking information sent to Unity via a UDP connection. Overall the system's latency from the head movement to the sound update was less than 30ms which is below the perceptual lag for binaural listening \cite{yairi2008influence, sankaran2016perceptual}. Listeners were placed in a room with a grid of angular markings on 3 sides of them at $10^\circ$ intervals for both azimuth and elevation. A white noise stimulus was played for a maximum of 5 seconds during which the listeners could make exploratory head movements within a maximum of $30^\circ$ of the forward facing angle. They were then asked to indicate their perceived source location by pointing at the best grid location. Experiments were conducted for sources in the front hemisphere and back hemisphere separately with sources coming from random locations in $\theta \in [-70^{\circ}, 70^\circ]$ and $\varphi \in [-30^\circ, 40^\circ]$. A brief calibration period was used where the listener could see the ground truth location for the first 4 examples while making the exploratory head movements. Each subject then evaluated 20 random locations for each candidate HRTF.

\begin{figure}[]
  \centering
  \includegraphics[width=1.0 \linewidth]{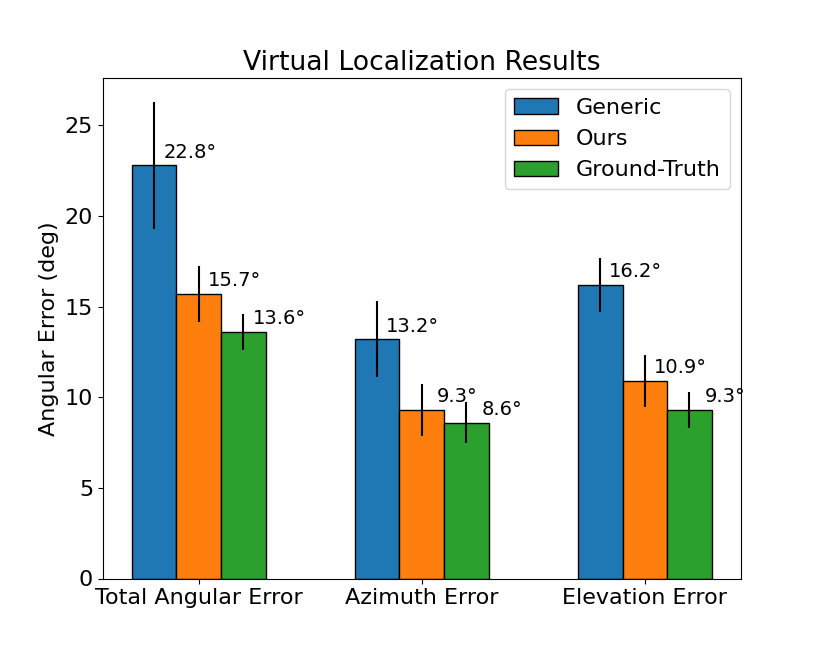}
  \caption{Localization results for the virtual auditory display experiment. Results are reported for 3 different experiments: a generic HRTF, the HRTF predicted using our method, and the ground-truth anechoic HRTF described in Section \ref{sec:anechoic}. For each experiment, we first show the total angle difference between the source and prediction. We then show the prediction error broken down by azimuth and elevation error. Results are averaged over all subjects and trials. Error bars shown are the first standard error of the mean.}
  \label{fig:localization}
\end{figure}

Results are reported in Figure \ref{fig:localization}. For both total angular error and elevation error, listeners performed significantly better with our method (p < 0.01) compared to a generic HRTF. In addition, the localization error with our method was close to that of the anechoic ground-truth HRTF. We note that, although the mean azimuth error was better with our method and the ground-truth HRTF compared to a generic HRTF, it was not statistically significant (p > 0.05). We hypothesize that this is because ITDs are the primary method used by humans for azimuth inference, and both the ground-truth HRTF and our method contained generic ITDs with only personalized spectral features. Statistical significance was computed with an independent-samples t test between the two candidate distributions.

\subsection{Front Back Confusion}
The last experiment conducted was a front-back confusion test using rendered sounds. A short white noise stimulus was rendered at a random location using a candidate HRTF and played back to the listener through headphones. The listener then had to predict whether the source was coming from the front or back hemisphere. The locations used were $\theta \in [-70^{\circ}, 70^\circ]$ in the front and back, and $\varphi \in [-30^\circ, 40^\circ]$. Like the previous experiment, the listener received the ground truth answer for the first 4 locations. However, unlike the previous experiment, the listener was not allowed to make exploratory head movements and had to predict front or back based on the rendering alone. Each subject then evaluated 30 random locations per HRTF before moving on to the next HRTF. The results are shown in table 2 which once again show a significant improvement (p < 0.01) when using our method compared to a generic HRTF.

\begin{table}\label{tab:frontback}
  \begin{tabular}{cc}
    \toprule
    Method & Front-back confusion rate \\
    \midrule
    Generic & 29.0\% $\pm$ 5.4 \\
    \textbf{Ours} & \textbf{14.8\% $\pm$ 4.6} \\
    GT HRTF & 9.6\% $\pm$ 4.2 \\
    \bottomrule
  \end{tabular}
  \vspace{10pt}
  \caption{Front-back confusion with rendered sounds. We report the percent of times the listeners made an error, along with the first standard deviation}
\end{table}

\section{Limitations and Conclusion}

Our method shows a strong ability to solve for a listener's HRTF using only binaural recordings of in-the-wild sounds and relative head tracking information. However, there are several limitations that need to be acknowledged.

First, our method was only demonstrated with a single stationary noise source. Such scenarios are limited in everyday settings, and solving for the HRTF with multiple sources or moving sources would present additional challenges. It would be necessary to localize moving sources and separate the contributions to the recording from multiple sources. Second, the user still has to actively localize the sources at the beginning of each recording, which presents an additional burden compared to a fully passive HRTF estimation method. This could be resolved by using a binaural localization method, and erroneous localizations could be compensated through outlier detection methods. Finally, the microphones in commercial earbuds are often not exactly at the ear canal entrance. The effect of the earbud on the HRTF would need to be taken into account through careful measurements of the earbud system. Despite these limitations, our method for HRTF estimation has immense potential as wireless earbuds proliferate among everyday users. We show strong performance on a variety of real-world user studies, and we hope that our method can be incorporated into commercial earbud systems in the near future.

\section{Acknowledgements}

This work was supported by the UW Reality Lab, Meta,
Google, OPPO, and Amazon.

\bibliography{references} 


\begin{thebibliography}{55}


\ifx \showCODEN    \undefined \def \showCODEN     #1{\unskip}     \fi
\ifx \showDOI      \undefined \def \showDOI       #1{#1}\fi
\ifx \showISBNx    \undefined \def \showISBNx     #1{\unskip}     \fi
\ifx \showISBNxiii \undefined \def \showISBNxiii  #1{\unskip}     \fi
\ifx \showISSN     \undefined \def \showISSN      #1{\unskip}     \fi
\ifx \showLCCN     \undefined \def \showLCCN      #1{\unskip}     \fi
\ifx \shownote     \undefined \def \shownote      #1{#1}          \fi
\ifx \showarticletitle \undefined \def \showarticletitle #1{#1}   \fi
\ifx \showURL      \undefined \def \showURL       {\relax}        \fi
\providecommand\bibfield[2]{#2}
\providecommand\bibinfo[2]{#2}
\providecommand\natexlab[1]{#1}
\providecommand\showeprint[2][]{arXiv:#2}

\bibitem[Algazi et~al\mbox{.}(2001)]%
        {algazi2001cipic}
\bibfield{author}{\bibinfo{person}{V~Ralph Algazi}, \bibinfo{person}{Richard~O
  Duda}, \bibinfo{person}{Dennis~M Thompson}, {and} \bibinfo{person}{Carlos
  Avendano}.} \bibinfo{year}{2001}\natexlab{}.
\newblock \showarticletitle{The cipic hrtf database}. In
  \bibinfo{booktitle}{\emph{Proceedings of the 2001 IEEE Workshop on the
  Applications of Signal Processing to Audio and Acoustics (Cat. No.
  01TH8575)}}. IEEE, \bibinfo{pages}{99--102}.
\newblock


\bibitem[Apple({[n.\,d.]})]%
        {applepersonalized}
\bibfield{author}{\bibinfo{person}{Apple}.}
  \bibinfo{year}{[n.\,d.]}\natexlab{}.
\newblock \bibinfo{booktitle}{\emph{Listen with Personalized Spatial Audio for
  AirPods and Beats}}.
\newblock
\urldef\tempurl%
\url{https://support.apple.com/en-us/HT213318}
\showURL{%
\tempurl}
\newblock
\shownote{Accessed on: June 1, 2023}.


\bibitem[Apple(2023a)]%
        {airpodspecs}
\bibfield{author}{\bibinfo{person}{Apple}.} \bibinfo{year}{2023}\natexlab{a}.
\newblock \bibinfo{booktitle}{\emph{AirPods (3rd generation)}}.
\newblock
\urldef\tempurl%
\url{https://www.apple.com/airpods-3rd-generation/specs/}
\showURL{%
\tempurl}
\newblock
\shownote{Accessed on: June 1, 2023}.


\bibitem[Apple(2023b)]%
        {airpodscmmotion}
\bibfield{author}{\bibinfo{person}{Apple}.} \bibinfo{year}{2023}\natexlab{b}.
\newblock \bibinfo{booktitle}{\emph{CMHeadphoneMotionManager}}.
\newblock
\urldef\tempurl%
\url{https://developer.apple.com/documentation/coremotion/cmheadphonemotionmanager}
\showURL{%
\tempurl}
\newblock
\shownote{Accessed on: June 1, 2023}.


\bibitem[Bazarevsky et~al\mbox{.}(2019)]%
        {bazarevsky2019blazeface}
\bibfield{author}{\bibinfo{person}{Valentin Bazarevsky}, \bibinfo{person}{Yury
  Kartynnik}, \bibinfo{person}{Andrey Vakunov}, \bibinfo{person}{Karthik
  Raveendran}, {and} \bibinfo{person}{Matthias Grundmann}.}
  \bibinfo{year}{2019}\natexlab{}.
\newblock \bibinfo{title}{BlazeFace: Sub-millisecond Neural Face Detection on
  Mobile GPUs}.
\newblock
\newblock
\showeprint[arxiv]{1907.05047}~[cs.CV]


\bibitem[ben hur et~al\mbox{.}(2020)]%
        {ben-hur2020localization}
\bibfield{author}{\bibinfo{person}{zamir ben hur}, \bibinfo{person}{david
  alon}, \bibinfo{person}{philip~w. robinson}, {and} \bibinfo{person}{ravish
  mehra}.} \bibinfo{year}{2020}\natexlab{}.
\newblock \showarticletitle{localization of virtual sounds in dynamic listening
  using sparse hrtfs}.
\newblock \bibinfo{journal}{\emph{journal of the audio engineering society}}
  (\bibinfo{date}{august} \bibinfo{year}{2020}).
\newblock


\bibitem[Chatterjee et~al\mbox{.}(2022)]%
        {Chatterjee_2022}
\bibfield{author}{\bibinfo{person}{Ishan Chatterjee}, \bibinfo{person}{Maruchi
  Kim}, \bibinfo{person}{Vivek Jayaram}, \bibinfo{person}{Shyamnath Gollakota},
  \bibinfo{person}{Ira Kemelmacher}, \bibinfo{person}{Shwetak Patel}, {and}
  \bibinfo{person}{Steven~M. Seitz}.} \bibinfo{year}{2022}\natexlab{}.
\newblock \showarticletitle{{ClearBuds}}. In
  \bibinfo{booktitle}{\emph{Proceedings of the 20th Annual International
  Conference on Mobile Systems, Applications and Services}}.
  \bibinfo{publisher}{{ACM}}.
\newblock
\urldef\tempurl%
\url{https://doi.org/10.1145/3498361.3538933}
\showDOI{\tempurl}


\bibitem[Chen et~al\mbox{.}(2019)]%
        {chen2019autoencoding}
\bibfield{author}{\bibinfo{person}{Tzu-Yu Chen}, \bibinfo{person}{Tzu-Hsuan
  Kuo}, {and} \bibinfo{person}{Tai-Shih Chi}.} \bibinfo{year}{2019}\natexlab{}.
\newblock \showarticletitle{Autoencoding HRTFs for DNN based HRTF
  personalization using anthropometric features}. In
  \bibinfo{booktitle}{\emph{ICASSP 2019-2019 IEEE International Conference on
  Acoustics, Speech and Signal Processing (ICASSP)}}. IEEE,
  \bibinfo{pages}{271--275}.
\newblock


\bibitem[Chun et~al\mbox{.}(2017)]%
        {chun2017deep}
\bibfield{author}{\bibinfo{person}{Chan~Jun Chun}, \bibinfo{person}{Jung~Min
  Moon}, \bibinfo{person}{Geon~Woo Lee}, \bibinfo{person}{Nam~Kyun Kim}, {and}
  \bibinfo{person}{Hong~Kook Kim}.} \bibinfo{year}{2017}\natexlab{}.
\newblock \showarticletitle{Deep neural network based HRTF personalization
  using anthropometric measurements}. In \bibinfo{booktitle}{\emph{Audio
  Engineering Society Convention 143}}. Audio Engineering Society.
\newblock


\bibitem[Diepold et~al\mbox{.}(2010)]%
        {diepold2010hrtf}
\bibfield{author}{\bibinfo{person}{Klaus Diepold}, \bibinfo{person}{Marko
  Durkovic}, {and} \bibinfo{person}{Florian Sagstetter}.}
  \bibinfo{year}{2010}\natexlab{}.
\newblock \showarticletitle{HRTF Measurements with Recorded Reference Signal}.
  In \bibinfo{booktitle}{\emph{Audio Engineering Society Convention 129}}.
  Audio Engineering Society.
\newblock


\bibitem[Gardner et~al\mbox{.}(1994)]%
        {gardner1994hrft}
\bibfield{author}{\bibinfo{person}{Bill Gardner}, \bibinfo{person}{Keith
  Martin}, {et~al\mbox{.}}} \bibinfo{year}{1994}\natexlab{}.
\newblock \showarticletitle{HRFT Measurements of a KEMAR Dummy-head
  Microphone}.
\newblock  (\bibinfo{year}{1994}).
\newblock


\bibitem[Gemmeke et~al\mbox{.}(2017)]%
        {audioset}
\bibfield{author}{\bibinfo{person}{Jort~F. Gemmeke}, \bibinfo{person}{Daniel
  P.~W. Ellis}, \bibinfo{person}{Dylan Freedman}, \bibinfo{person}{Aren
  Jansen}, \bibinfo{person}{Wade Lawrence}, \bibinfo{person}{R.~Channing
  Moore}, \bibinfo{person}{Manoj Plakal}, {and} \bibinfo{person}{Marvin
  Ritter}.} \bibinfo{year}{2017}\natexlab{}.
\newblock \showarticletitle{Audio Set: An ontology and human-labeled dataset
  for audio events}. In \bibinfo{booktitle}{\emph{Proc. IEEE ICASSP 2017}}.
  \bibinfo{address}{New Orleans, LA}.
\newblock


\bibitem[Google(2023)]%
        {googlemediapipe}
\bibfield{author}{\bibinfo{person}{Google}.} \bibinfo{year}{2023}\natexlab{}.
\newblock \bibinfo{title}{Mediapipe}.
\newblock
\newblock
\urldef\tempurl%
\url{https://github.com/google/mediapipe}
\showURL{%
\tempurl}
\newblock
\shownote{Accessed on: June 1, 2023}.


\bibitem[Grishchenko et~al\mbox{.}(2020)]%
        {grishchenko2020attention}
\bibfield{author}{\bibinfo{person}{Ivan Grishchenko}, \bibinfo{person}{Artsiom
  Ablavatski}, \bibinfo{person}{Yury Kartynnik}, \bibinfo{person}{Karthik
  Raveendran}, {and} \bibinfo{person}{Matthias Grundmann}.}
  \bibinfo{year}{2020}\natexlab{}.
\newblock \bibinfo{title}{Attention Mesh: High-fidelity Face Mesh Prediction in
  Real-time}.
\newblock
\newblock
\showeprint[arxiv]{2006.10962}~[cs.CV]


\bibitem[Hu et~al\mbox{.}(2006)]%
        {hu2006head}
\bibfield{author}{\bibinfo{person}{Hongmei Hu}, \bibinfo{person}{Lin Zhou},
  \bibinfo{person}{Jie Zhang}, \bibinfo{person}{Hao Ma}, {and}
  \bibinfo{person}{Zhenyang Wu}.} \bibinfo{year}{2006}\natexlab{}.
\newblock \showarticletitle{Head related transfer function personalization
  based on multiple regression analysis}. In \bibinfo{booktitle}{\emph{2006
  International conference on computational intelligence and security}},
  Vol.~\bibinfo{volume}{2}. IEEE, \bibinfo{pages}{1829--1832}.
\newblock


\bibitem[Huttunen et~al\mbox{.}(2007)]%
        {huttunen2007simulation}
\bibfield{author}{\bibinfo{person}{Tomi Huttunen}, \bibinfo{person}{Eira~T
  Sepp{\"a}l{\"a}}, \bibinfo{person}{Ole Kirkeby}, \bibinfo{person}{Asta
  K{\"a}rkk{\"a}inen}, {and} \bibinfo{person}{Leo K{\"a}rkk{\"a}inen}.}
  \bibinfo{year}{2007}\natexlab{}.
\newblock \showarticletitle{Simulation of the transfer function for a
  head-and-torso model over the entire audible frequency range}.
\newblock \bibinfo{journal}{\emph{Journal of Computational Acoustics}}
  \bibinfo{volume}{15}, \bibinfo{number}{04} (\bibinfo{year}{2007}),
  \bibinfo{pages}{429--448}.
\newblock


\bibitem[Huttunen et~al\mbox{.}(2014)]%
        {huttunen2014rapid}
\bibfield{author}{\bibinfo{person}{Tomi Huttunen}, \bibinfo{person}{Antti
  Vanne}, \bibinfo{person}{Stine Harder}, \bibinfo{person}{Rasmus~Reinhold
  Paulsen}, \bibinfo{person}{Sam King}, \bibinfo{person}{Lee Perry-Smith},
  {and} \bibinfo{person}{Leo K{\"a}rkk{\"a}inen}.}
  \bibinfo{year}{2014}\natexlab{}.
\newblock \showarticletitle{Rapid generation of personalized HRTFs}. In
  \bibinfo{booktitle}{\emph{Audio Engineering Society Conference: 55th
  International Conference: Spatial Audio}}. Audio Engineering Society.
\newblock


\bibitem[Ito et~al\mbox{.}(2022)]%
        {ito2022headrelated}
\bibfield{author}{\bibinfo{person}{Yuki Ito}, \bibinfo{person}{Tomohiko
  Nakamura}, \bibinfo{person}{Shoichi Koyama}, {and} \bibinfo{person}{Hiroshi
  Saruwatari}.} \bibinfo{year}{2022}\natexlab{}.
\newblock \bibinfo{title}{Head-Related Transfer Function Interpolation from
  Spatially Sparse Measurements Using Autoencoder with Source Position
  Conditioning}.
\newblock
\newblock
\showeprint[arxiv]{2207.10967}~[cs.SD]


\bibitem[Jenrungrot et~al\mbox{.}(2020)]%
        {jenrungrot2020cone}
\bibfield{author}{\bibinfo{person}{Teerapat Jenrungrot}, \bibinfo{person}{Vivek
  Jayaram}, \bibinfo{person}{Steve Seitz}, {and} \bibinfo{person}{Ira
  Kemelmacher-Shlizerman}.} \bibinfo{year}{2020}\natexlab{}.
\newblock \showarticletitle{The cone of silence: Speech separation by
  localization}.
\newblock \bibinfo{journal}{\emph{Advances in Neural Information Processing
  Systems}}  \bibinfo{volume}{33} (\bibinfo{year}{2020}),
  \bibinfo{pages}{20925--20938}.
\newblock


\bibitem[Langendijk and Bronkhorst(2000)]%
        {langendijk}
\bibfield{author}{\bibinfo{person}{Erno Langendijk} {and}
  \bibinfo{person}{Adelbert Bronkhorst}.} \bibinfo{year}{2000}\natexlab{}.
\newblock \showarticletitle{Fidelity of three-dimensional-sound reproduction
  using a virtual auditory display}.
\newblock \bibinfo{journal}{\emph{The Journal of the Acoustical Society of
  America}}  \bibinfo{volume}{107} (\bibinfo{date}{02} \bibinfo{year}{2000}),
  \bibinfo{pages}{528--37}.
\newblock
\urldef\tempurl%
\url{https://doi.org/10.1121/1.428321}
\showDOI{\tempurl}


\bibitem[Li and Peissig(2017)]%
        {fast2dheadmovements}
\bibfield{author}{\bibinfo{person}{Song Li} {and} \bibinfo{person}{Jürgen
  Peissig}.} \bibinfo{year}{2017}\natexlab{}.
\newblock \showarticletitle{Fast estimation of 2D individual HRTFs with
  arbitrary head movements}. In \bibinfo{booktitle}{\emph{2017 22nd
  International Conference on Digital Signal Processing (DSP)}}.
  \bibinfo{pages}{1--5}.
\newblock
\urldef\tempurl%
\url{https://doi.org/10.1109/ICDSP.2017.8096086}
\showDOI{\tempurl}


\bibitem[Logitech(2023)]%
        {logitechpersonalized}
\bibfield{author}{\bibinfo{person}{Logitech}.} \bibinfo{year}{2023}\natexlab{}.
\newblock \bibinfo{booktitle}{\emph{Personalized Spatial Audio with Head
  Tracking}}.
\newblock
\urldef\tempurl%
\url{https://embody.co/pages/gaming-logitech}
\showURL{%
\tempurl}
\newblock
\shownote{Accessed on: June 1, 2023}.


\bibitem[Majdak et~al\mbox{.}(2013)]%
        {majdak2013spatially}
\bibfield{author}{\bibinfo{person}{Piotr Majdak}, \bibinfo{person}{Yukio
  Iwaya}, \bibinfo{person}{Thibaut Carpentier}, \bibinfo{person}{Rozenn Nicol},
  \bibinfo{person}{Matthieu Parmentier}, \bibinfo{person}{Agnieszka Roginska},
  \bibinfo{person}{Y{\^o}iti Suzuki}, \bibinfo{person}{Kankji Watanabe},
  \bibinfo{person}{Hagen Wierstorf}, \bibinfo{person}{Harald Ziegelwanger},
  {et~al\mbox{.}}} \bibinfo{year}{2013}\natexlab{}.
\newblock \showarticletitle{Spatially oriented format for acoustics: A data
  exchange format representing head-related transfer functions}. In
  \bibinfo{booktitle}{\emph{Audio Engineering Society Convention 134}}. Audio
  Engineering Society.
\newblock


\bibitem[Makous and Middlebrooks(1990)]%
        {makous1990two}
\bibfield{author}{\bibinfo{person}{James~C Makous} {and}
  \bibinfo{person}{John~C Middlebrooks}.} \bibinfo{year}{1990}\natexlab{}.
\newblock \showarticletitle{Two-dimensional sound localization by human
  listeners}.
\newblock \bibinfo{journal}{\emph{The journal of the Acoustical Society of
  America}} \bibinfo{volume}{87}, \bibinfo{number}{5} (\bibinfo{year}{1990}),
  \bibinfo{pages}{2188--2200}.
\newblock


\bibitem[Meshram et~al\mbox{.}(2014a)]%
        {mesharmmesh}
\bibfield{author}{\bibinfo{person}{A. Meshram}, \bibinfo{person}{Ravish Mehra},
  {and} \bibinfo{person}{Dinesh Manocha}.} \bibinfo{year}{2014}\natexlab{a}.
\newblock \showarticletitle{Efficient HRTF computation using adaptive
  rectangular decomposition}.
\newblock \bibinfo{journal}{\emph{Proceedings of the AES International
  Conference}}  \bibinfo{volume}{2014} (\bibinfo{date}{01}
  \bibinfo{year}{2014}).
\newblock


\bibitem[Meshram et~al\mbox{.}(2014b)]%
        {pHRTF}
\bibfield{author}{\bibinfo{person}{Alok Meshram}, \bibinfo{person}{Ravish
  Mehra}, \bibinfo{person}{Hongsheng Yang}, \bibinfo{person}{Enrique Dunn},
  \bibinfo{person}{Jan-Michael Franm}, {and} \bibinfo{person}{Dinesh Manocha}.}
  \bibinfo{year}{2014}\natexlab{b}.
\newblock \showarticletitle{P-HRTF: Efficient personalized HRTF computation for
  high-fidelity spatial sound}. In \bibinfo{booktitle}{\emph{2014 IEEE
  International Symposium on Mixed and Augmented Reality (ISMAR)}}.
  \bibinfo{pages}{53--61}.
\newblock
\urldef\tempurl%
\url{https://doi.org/10.1109/ISMAR.2014.6948409}
\showDOI{\tempurl}


\bibitem[Middlebrooks(1999)]%
        {middlebrooks1999individual}
\bibfield{author}{\bibinfo{person}{John~C Middlebrooks}.}
  \bibinfo{year}{1999}\natexlab{}.
\newblock \showarticletitle{Individual differences in external-ear transfer
  functions reduced by scaling in frequency}.
\newblock \bibinfo{journal}{\emph{The Journal of the Acoustical Society of
  America}} \bibinfo{volume}{106}, \bibinfo{number}{3} (\bibinfo{year}{1999}),
  \bibinfo{pages}{1480--1492}.
\newblock


\bibitem[Mohan et~al\mbox{.}(2003)]%
        {mohancomputervision}
\bibfield{author}{\bibinfo{person}{A. Mohan}, \bibinfo{person}{R. Duraiswami},
  \bibinfo{person}{D.N. Zotkin}, \bibinfo{person}{D. DeMenthon}, {and}
  \bibinfo{person}{L.S. Davis}.} \bibinfo{year}{2003}\natexlab{}.
\newblock \showarticletitle{Using computer vision to generate customized
  spatial audio}. In \bibinfo{booktitle}{\emph{2003 International Conference on
  Multimedia and Expo. ICME '03. Proceedings (Cat. No.03TH8698)}},
  Vol.~\bibinfo{volume}{3}. \bibinfo{pages}{III--57}.
\newblock
\urldef\tempurl%
\url{https://doi.org/10.1109/ICME.2003.1221247}
\showDOI{\tempurl}


\bibitem[M{\o}ller et~al\mbox{.}(1995)]%
        {moller1995head}
\bibfield{author}{\bibinfo{person}{Henrik M{\o}ller},
  \bibinfo{person}{Michael~Friis S{\o}rensen}, \bibinfo{person}{Dorte
  Hammersh{\o}i}, {and} \bibinfo{person}{Clemen~Boje Jensen}.}
  \bibinfo{year}{1995}\natexlab{}.
\newblock \showarticletitle{Head-related transfer functions of human subjects}.
\newblock \bibinfo{journal}{\emph{Journal of the Audio Engineering Society}}
  \bibinfo{volume}{43}, \bibinfo{number}{5} (\bibinfo{year}{1995}),
  \bibinfo{pages}{300--321}.
\newblock


\bibitem[Peterson(2021)]%
        {airpodssales}
\bibfield{author}{\bibinfo{person}{Mike Peterson}.}
  \bibinfo{year}{2021}\natexlab{}.
\newblock \bibinfo{title}{Apple AirPods, Beats dominated audio wearable market
  in 2020}.
\newblock
\newblock
\urldef\tempurl%
\url{https://appleinsider.com/articles/21/03/30/apple-airpods-beats-dominated-audio-wearable-market-in-2020}
\showURL{%
\tempurl}
\newblock
\shownote{Accessed on: June 1, 2023}.


\bibitem[Professionals(2022)]%
        {binauralmics}
\bibfield{author}{\bibinfo{person}{Sound Professionals}.}
  \bibinfo{year}{2022}\natexlab{}.
\newblock \bibinfo{title}{SP-TFB-2 – Low noise in-ear Binaural microphones}.
\newblock
\newblock
\urldef\tempurl%
\url{https://soundprofessionals.com/product/SP-TFB-2/}
\showURL{%
\tempurl}
\newblock
\shownote{Accessed on: June 1, 2023}.


\bibitem[Reijniers et~al\mbox{.}(2017)]%
        {reijniers2017diy}
\bibfield{author}{\bibinfo{person}{Jonas Reijniers}, \bibinfo{person}{Bart
  Partoens}, {and} \bibinfo{person}{Herbert Peremans}.}
  \bibinfo{year}{2017}\natexlab{}.
\newblock \showarticletitle{DIY Measurement of your Personal Hrtf at Home:
  Low-Cost, Fast and Validated}.
\newblock \bibinfo{journal}{\emph{journal of the audio engineering society}}
  (\bibinfo{date}{october} \bibinfo{year}{2017}).
\newblock


\bibitem[Reijniers et~al\mbox{.}(2020)]%
        {headmovements}
\bibfield{author}{\bibinfo{person}{Jonas Reijniers}, \bibinfo{person}{Bart
  Partoens}, \bibinfo{person}{Jan Steckel}, {and} \bibinfo{person}{Herbert
  Peremans}.} \bibinfo{year}{2020}\natexlab{}.
\newblock \showarticletitle{HRTF Measurement by Means of Unsupervised Head
  Movements With Respect to a Single Fixed Speaker}.
\newblock \bibinfo{journal}{\emph{IEEE Access}}  \bibinfo{volume}{8}
  (\bibinfo{year}{2020}), \bibinfo{pages}{92287--92300}.
\newblock
\urldef\tempurl%
\url{https://doi.org/10.1109/ACCESS.2020.2994932}
\showDOI{\tempurl}


\bibitem[Richard et~al\mbox{.}(2022)]%
        {richard2022deep}
\bibfield{author}{\bibinfo{person}{Alexander Richard}, \bibinfo{person}{Peter
  Dodds}, {and} \bibinfo{person}{Vamsi~Krishna Ithapu}.}
  \bibinfo{year}{2022}\natexlab{}.
\newblock \bibinfo{title}{Deep Impulse Responses: Estimating and Parameterizing
  Filters with Deep Networks}.
\newblock
\newblock
\showeprint[arxiv]{2202.03416}~[cs.SD]


\bibitem[Ronneberger et~al\mbox{.}(2015)]%
        {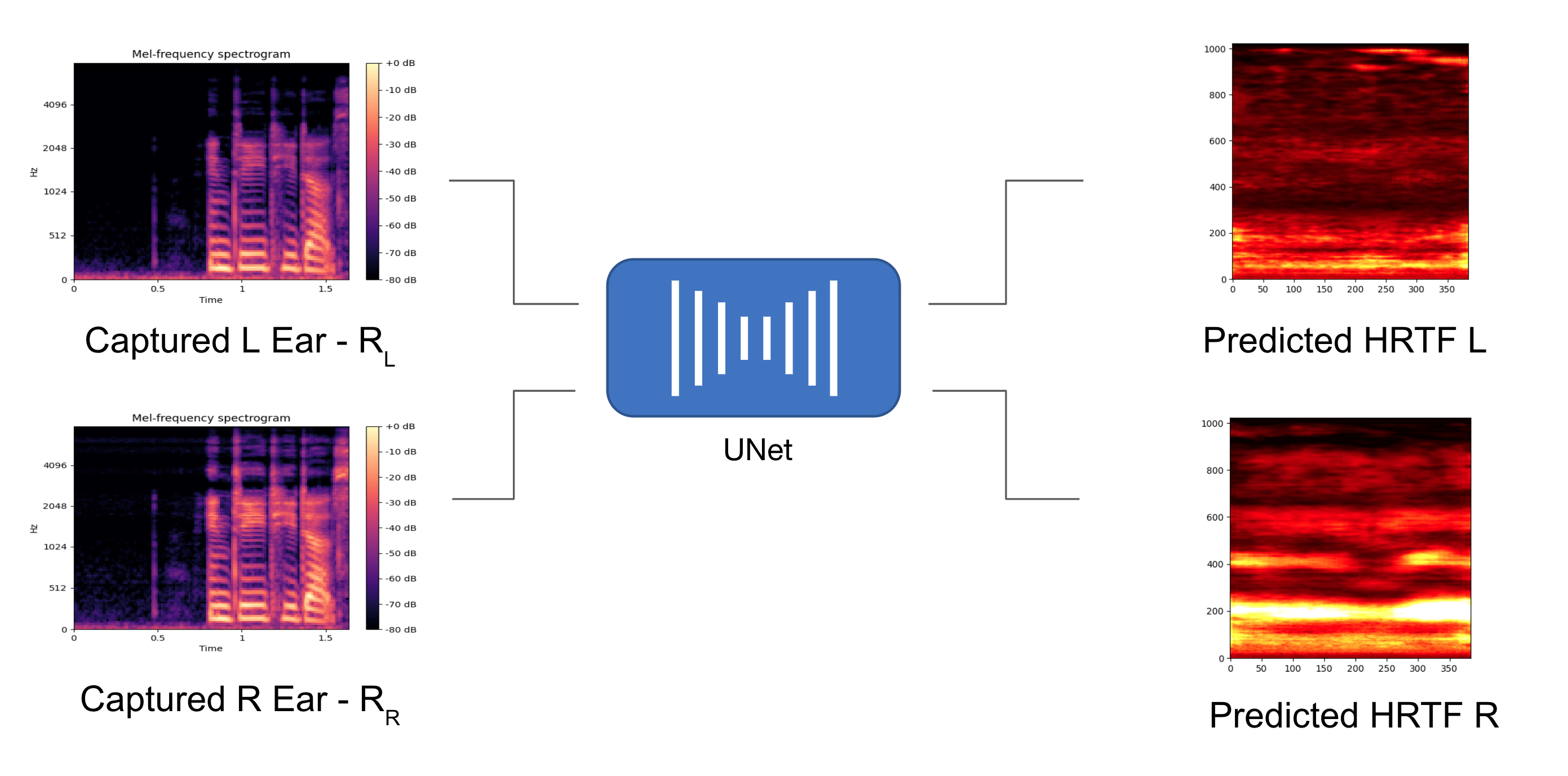}
\bibfield{author}{\bibinfo{person}{Olaf Ronneberger}, \bibinfo{person}{Philipp
  Fischer}, {and} \bibinfo{person}{Thomas Brox}.}
  \bibinfo{year}{2015}\natexlab{}.
\newblock \showarticletitle{U-net: Convolutional networks for biomedical image
  segmentation}. In \bibinfo{booktitle}{\emph{International Conference on
  Medical image computing and computer-assisted intervention}}. Springer,
  \bibinfo{pages}{234--241}.
\newblock


\bibitem[Sankaran et~al\mbox{.}(2016)]%
        {sankaran2016perceptual}
\bibfield{author}{\bibinfo{person}{Narayan Sankaran}, \bibinfo{person}{James
  Hillis}, \bibinfo{person}{Marina Zannoli}, {and} \bibinfo{person}{Ravish
  Mehra}.} \bibinfo{year}{2016}\natexlab{}.
\newblock \showarticletitle{Perceptual thresholds of spatial audio update
  latency in virtual auditory and audiovisual environments}.
\newblock \bibinfo{journal}{\emph{The Journal of the Acoustical Society of
  America}} \bibinfo{volume}{140}, \bibinfo{number}{4} (\bibinfo{year}{2016}),
  \bibinfo{pages}{3008--3008}.
\newblock


\bibitem[Schoon(2023)]%
        {samsungbuds}
\bibfield{author}{\bibinfo{person}{Ben Schoon}.}
  \bibinfo{year}{2023}\natexlab{}.
\newblock \bibinfo{booktitle}{\emph{Samsung Galaxy Buds 2 Pro can now record
  360-degree binaural audio for videos from your phone}}.
\newblock
\urldef\tempurl%
\url{https://9to5google.com/2023/01/12/samsung-buds-binaural-audio-recording}
\showURL{%
\tempurl}
\newblock
\shownote{Accessed on: June 1, 2023}.


\bibitem[Seib et~al\mbox{.}(2020)]%
        {seib2020mixing}
\bibfield{author}{\bibinfo{person}{Viktor Seib}, \bibinfo{person}{Benjamin
  Lange}, {and} \bibinfo{person}{Stefan Wirtz}.}
  \bibinfo{year}{2020}\natexlab{}.
\newblock \bibinfo{title}{Mixing Real and Synthetic Data to Enhance Neural
  Network Training -- A Review of Current Approaches}.
\newblock
\newblock
\showeprint[arxiv]{2007.08781}~[cs.CV]


\bibitem[Sony(2023)]%
        {sonypersonalized}
\bibfield{author}{\bibinfo{person}{Sony}.} \bibinfo{year}{2023}\natexlab{}.
\newblock \bibinfo{booktitle}{\emph{360 Reality Audio}}.
\newblock
\urldef\tempurl%
\url{https://electronics.sony.com/360-reality-audio}
\showURL{%
\tempurl}
\newblock
\shownote{Accessed on: June 1, 2023}.


\bibitem[Southampton({[n.\,d.]})]%
        {southampton}
\bibfield{author}{\bibinfo{person}{University~of Southampton}.}
  \bibinfo{year}{[n.\,d.]}\natexlab{}.
\newblock \bibinfo{title}{HRTF Mesaurement System}.
\newblock
\newblock
\urldef\tempurl%
\url{https://resource.isvr.soton.ac.uk/FDAG/VAP/html/facilities.html}
\showURL{%
\tempurl}


\bibitem[Sridhar and Choueiri(2017)]%
        {sridhar2017method}
\bibfield{author}{\bibinfo{person}{Rahulram Sridhar} {and}
  \bibinfo{person}{Edgar~Y Choueiri}.} \bibinfo{year}{2017}\natexlab{}.
\newblock \showarticletitle{A method for efficiently calculating head-related
  transfer functions directly from head scan point clouds}. In
  \bibinfo{booktitle}{\emph{143rd Audio Engineering Society Convention 2017}}.
\newblock


\bibitem[Sridhar et~al\mbox{.}(2017)]%
        {sridhar2017database}
\bibfield{author}{\bibinfo{person}{Rahulram Sridhar}, \bibinfo{person}{Joseph~G
  Tylka}, {and} \bibinfo{person}{Edgar Choueiri}.}
  \bibinfo{year}{2017}\natexlab{}.
\newblock \showarticletitle{A database of head-related transfer functions and
  morphological measurements}. In \bibinfo{booktitle}{\emph{Audio Engineering
  Society Convention 143}}. Audio Engineering Society.
\newblock


\bibitem[Steinmetz et~al\mbox{.}(2021)]%
        {filteredNoiseRIR}
\bibfield{author}{\bibinfo{person}{Christian~J. Steinmetz},
  \bibinfo{person}{Vamsi~Krishna Ithapu}, {and} \bibinfo{person}{Paul
  Calamia}.} \bibinfo{year}{2021}\natexlab{}.
\newblock \showarticletitle{Filtered Noise Shaping for Time Domain Room Impulse
  Response Estimation from Reverberant Speech}. In
  \bibinfo{booktitle}{\emph{2021 IEEE Workshop on Applications of Signal
  Processing to Audio and Acoustics (WASPAA)}}. \bibinfo{pages}{221--225}.
\newblock
\urldef\tempurl%
\url{https://doi.org/10.1109/WASPAA52581.2021.9632680}
\showDOI{\tempurl}


\bibitem[Watanabe et~al\mbox{.}(2014)]%
        {watanabe2014dataset}
\bibfield{author}{\bibinfo{person}{Kanji Watanabe}, \bibinfo{person}{Yukio
  Iwaya}, \bibinfo{person}{Y{\^o}iti Suzuki}, \bibinfo{person}{Shouichi
  Takane}, {and} \bibinfo{person}{Sojun Sato}.}
  \bibinfo{year}{2014}\natexlab{}.
\newblock \showarticletitle{Dataset of head-related transfer functions measured
  with a circular loudspeaker array}.
\newblock \bibinfo{journal}{\emph{Acoustical science and technology}}
  \bibinfo{volume}{35}, \bibinfo{number}{3} (\bibinfo{year}{2014}),
  \bibinfo{pages}{159--165}.
\newblock


\bibitem[Wenzel et~al\mbox{.}(1993)]%
        {wenzel1993localization}
\bibfield{author}{\bibinfo{person}{Elizabeth~M Wenzel},
  \bibinfo{person}{Marianne Arruda}, \bibinfo{person}{Doris~J Kistler}, {and}
  \bibinfo{person}{Frederic~L Wightman}.} \bibinfo{year}{1993}\natexlab{}.
\newblock \showarticletitle{Localization using nonindividualized head-related
  transfer functions}.
\newblock \bibinfo{journal}{\emph{The Journal of the Acoustical Society of
  America}} \bibinfo{volume}{94}, \bibinfo{number}{1} (\bibinfo{year}{1993}),
  \bibinfo{pages}{111--123}.
\newblock


\bibitem[Wichern et~al\mbox{.}(2019)]%
        {wham}
\bibfield{author}{\bibinfo{person}{Gordon Wichern}, \bibinfo{person}{Joe
  Antognini}, \bibinfo{person}{Michael Flynn}, \bibinfo{person}{Licheng~Richard
  Zhu}, \bibinfo{person}{Emmett McQuinn}, \bibinfo{person}{Dwight Crow},
  \bibinfo{person}{Ethan Manilow}, {and} \bibinfo{person}{Jonathan~Le Roux}.}
  \bibinfo{year}{2019}\natexlab{}.
\newblock \showarticletitle{WHAM!: Extending speech separation to noisy
  environments}.
\newblock \bibinfo{journal}{\emph{arXiv preprint arXiv:1907.01160}}
  (\bibinfo{year}{2019}).
\newblock


\bibitem[Yairi et~al\mbox{.}(2008)]%
        {yairi2008influence}
\bibfield{author}{\bibinfo{person}{Satoshi Yairi}, \bibinfo{person}{Yukio
  Iwaya}, {and} \bibinfo{person}{Y{\^o}iti Suzuki}.}
  \bibinfo{year}{2008}\natexlab{}.
\newblock \showarticletitle{Influence of large system latency of virtual
  auditory display on behavior of head movement in sound localization task}.
\newblock \bibinfo{journal}{\emph{Acta Acustica united with Acustica}}
  \bibinfo{volume}{94}, \bibinfo{number}{6} (\bibinfo{year}{2008}),
  \bibinfo{pages}{1016--1023}.
\newblock


\bibitem[Yamamoto and Igarashi(2017)]%
        {yamamoto2017fully}
\bibfield{author}{\bibinfo{person}{Kazuhiko Yamamoto} {and}
  \bibinfo{person}{Takeo Igarashi}.} \bibinfo{year}{2017}\natexlab{}.
\newblock \showarticletitle{Fully perceptual-based 3D spatial sound
  individualization with an adaptive variational autoencoder}.
\newblock \bibinfo{journal}{\emph{ACM Transactions on Graphics (TOG)}}
  \bibinfo{volume}{36}, \bibinfo{number}{6} (\bibinfo{year}{2017}),
  \bibinfo{pages}{1--13}.
\newblock


\bibitem[Yang and Choudhury(2021)]%
        {earables}
\bibfield{author}{\bibinfo{person}{Zhijian Yang} {and}
  \bibinfo{person}{Romit~Roy Choudhury}.} \bibinfo{year}{2021}\natexlab{}.
\newblock \showarticletitle{Personalizing Head Related Transfer Functions for
  Earables}. In \bibinfo{booktitle}{\emph{Proceedings of the 2021 ACM SIGCOMM
  2021 Conference}} (Virtual Event, USA) \emph{(\bibinfo{series}{SIGCOMM
  '21})}. \bibinfo{publisher}{Association for Computing Machinery},
  \bibinfo{address}{New York, NY, USA}, \bibinfo{pages}{137–150}.
\newblock
\showISBNx{9781450383837}
\urldef\tempurl%
\url{https://doi.org/10.1145/3452296.3472907}
\showDOI{\tempurl}


\bibitem[Zandi et~al\mbox{.}(2022)]%
        {zandi2022individualizing}
\bibfield{author}{\bibinfo{person}{Navid~H Zandi}, \bibinfo{person}{Awny~M
  El-Mohandes}, {and} \bibinfo{person}{Rong Zheng}.}
  \bibinfo{year}{2022}\natexlab{}.
\newblock \showarticletitle{Individualizing Head-Related Transfer Functions for
  Binaural Acoustic Applications}. In \bibinfo{booktitle}{\emph{2022 21st
  ACM/IEEE International Conference on Information Processing in Sensor
  Networks (IPSN)}}. IEEE, \bibinfo{pages}{105--117}.
\newblock


\bibitem[Zhao et~al\mbox{.}(2022)]%
        {zhao2022magnitude}
\bibfield{author}{\bibinfo{person}{Manlin Zhao}, \bibinfo{person}{Zhichao
  Sheng}, {and} \bibinfo{person}{Yong Fang}.} \bibinfo{year}{2022}\natexlab{}.
\newblock \showarticletitle{Magnitude modeling of personalized HRTF based on
  ear images and anthropometric measurements}.
\newblock \bibinfo{journal}{\emph{Applied Sciences}} \bibinfo{volume}{12},
  \bibinfo{number}{16} (\bibinfo{year}{2022}), \bibinfo{pages}{8155}.
\newblock


\bibitem[Zhi et~al\mbox{.}(2022)]%
        {zhiphoto}
\bibfield{author}{\bibinfo{person}{Bowen Zhi}, \bibinfo{person}{Dmitry~N.
  Zotkin}, {and} \bibinfo{person}{Ramani Duraiswami}.}
  \bibinfo{year}{2022}\natexlab{}.
\newblock \showarticletitle{Towards Fast And Convenient End-To-End HRTF
  Personalization}. In \bibinfo{booktitle}{\emph{ICASSP 2022 - 2022 IEEE
  International Conference on Acoustics, Speech and Signal Processing
  (ICASSP)}}. \bibinfo{pages}{441--445}.
\newblock
\urldef\tempurl%
\url{https://doi.org/10.1109/ICASSP43922.2022.9746315}
\showDOI{\tempurl}


\bibitem[Ziegelwanger et~al\mbox{.}(2015)]%
        {ziegelwanger2015mesh2hrtf}
\bibfield{author}{\bibinfo{person}{Harald Ziegelwanger},
  \bibinfo{person}{Wolfgang Kreuzer}, {and} \bibinfo{person}{Piotr Majdak}.}
  \bibinfo{year}{2015}\natexlab{}.
\newblock \showarticletitle{Mesh2hrtf: Open-source software package for the
  numerical calculation of head-related transfer functions}. In
  \bibinfo{booktitle}{\emph{22nd International Congress on Sound and
  Vibration}}.
\newblock


\bibitem[Zotkin et~al\mbox{.}(2003)]%
        {anthropometric}
\bibfield{author}{\bibinfo{person}{D.Y.N. Zotkin}, \bibinfo{person}{J. Hwang},
  \bibinfo{person}{R. Duraiswaini}, {and} \bibinfo{person}{L.S. Davis}.}
  \bibinfo{year}{2003}\natexlab{}.
\newblock \showarticletitle{HRTF personalization using anthropometric
  measurements}. In \bibinfo{booktitle}{\emph{2003 IEEE Workshop on
  Applications of Signal Processing to Audio and Acoustics (IEEE Cat.
  No.03TH8684)}}. \bibinfo{pages}{157--160}.
\newblock
\urldef\tempurl%
\url{https://doi.org/10.1109/ASPAA.2003.1285855}
\showDOI{\tempurl}


\bibitem[Zotkin et~al\mbox{.}(2002)]%
        {zotkin2002customizable}
\bibfield{author}{\bibinfo{person}{Dmitry~N Zotkin}, \bibinfo{person}{Ramani
  Duraiswami}, {and} \bibinfo{person}{Larry~S Davis}.}
  \bibinfo{year}{2002}\natexlab{}.
\newblock \showarticletitle{Customizable auditory displays}. Georgia Institute
  of Technology.
\newblock


\end{thebibliography}

\end{document}